\pdfoutput=1
\PassOptionsToPackage{protrusion, expansion}{microtype} %

\documentclass[sigconf,nonacm]{acmart}
\setcopyright{none}

\renewcommand\footnotetextcopyrightpermission[1]{}

\usepackage{array}
\usepackage{algorithm}
\usepackage{algorithmic}
\usepackage{multirow}
\usepackage[tight,footnotesize]{subfigure}
\definecolor{B}    {HTML}{2b66d3}   %
\definecolor{B2}   {HTML}{003399}   %
\definecolor{Bv}   {HTML}{0000EB}   %
\definecolor{R}    {HTML}{c9171e}
\definecolor{R2}   {HTML}{d7003a}
\definecolor{INK}  {HTML}{595857}
\definecolor{Y}    {HTML}{f1c40f}
\definecolor{G}    {HTML}{009a00}
\definecolor{GRAY} {HTML}{808080}
\definecolor{MAUVE}{HTML}{9400D1}

\usepackage{url, hyperref}                  %

\usepackage{amssymb}
\usepackage{amsmath}
\usepackage{color}
\usepackage{verbatim}
\usepackage{enumitem}
\usepackage{tikz}
\usepackage{listings}
\usepackage{graphicx}
\usepackage{balance}

\newcommand{\FloatBodyStyle}{\centering\footnotesize\fontfamily{phv}\selectfont\renewcommand{\arraystretch}{1.3}}

\definecolor{inferno_d1}{RGB}{ 51,  9, 95}
\definecolor{inferno_d2}{RGB}{120, 28,109}
\definecolor{inferno_d3}{RGB}{188, 54, 84}
\definecolor{inferno_d4}{RGB}{237,105, 36}
\definecolor{inferno_d5}{RGB}{251,182, 27}

\setlist{
  ,nosep
  ,itemindent=0pt
  ,leftmargin=*
  ,before=\vspace{.25\baselineskip}
  ,after=\vspace{.25\baselineskip}
}

\newcommand{\cusz}{{\scshape cuSZ}}
\newcommand{\thiswork}{\cusz-Hi}

\usepackage{lmodern}
\usepackage{helvet}

\hypersetup{colorlinks=true,allcolors=RoyalBlue
}

\begin{document}

\title[\cusz-Hi]
{Boosting Scientific Error-Bounded Lossy Compression through Optimized Synergistic Lossy-Lossless Orchestration}

\settopmatter{authorsperrow=4}

\author{Shixun Wu}
\authornote{Both authors contributed equally to this research.}
\affiliation{
  \institution{\makebox[10em][c]{Univ of California Riverside}}
  \city{Riverside}\state{CA}
  \country{USA}
}
\email{swu264@ucr.edu}

\author{Jinwen Pan}
\authornotemark[1]
\affiliation{
  \institution{\makebox[10em][c]{Technical Univ of Munich}}
  \city{Munich}\state{BY}
  \country{Germany}
}
\email{jinwen.pan@tum.de}

\author{Jinyang Liu}
\authornote{Corresponding author: Jinyang Liu, Department of Computer Science, University of Houston, Houston, TX 77204.}
\affiliation{%
  \institution{University of Houston}
  \city{Houston}\state{TX}
  \country{USA}
}
\email{jliu217@central.uh.edu}

\author{Jiannan Tian}
\affiliation{
  \institution{Oakland University}
  \city{Rochester}\state{MI}
  \country{USA}
}
\email{jtian@oakland.edu}

\author{Ziwei Qiu}
\affiliation{%
  \institution{University of Houston}
  \city{Houston}\state{TX}
  \country{USA}
}
\email{zqiu4@cougarnet.uh.edu}

\author{Jiajun Huang}
\affiliation{
  \institution{\makebox[10em][c]{University of South Florida}}
  \city{Tampa}\state{FL}
  \country{USA}
}
\email{jiajunhuang@usf.edu}

\author{Kai Zhao}
\affiliation{%
  \institution{Florida State University}
  \city{Tallahassee}\state{FL}
  \country{USA}
}
\email{kai.zhao@fsu.edu}

\author{Xin Liang}
\affiliation{%
  \institution{University of Kentucky}
  \city{Lexington}\state{KY}
  \country{USA}
}
\email{xliang@uky.edu}

\author{Sheng Di}
\affiliation{%
  \institution{Argonne National Lab}
  \city{Lemont}\state{IL}
  \country{USA}
}
\email{sdi1@anl.gov}

\author{Zizhong Chen}
\affiliation{
  \institution{\makebox[10em][c]{Univ of California Riverside}}
  \city{Riverside}\state{CA}
  \country{USA}
}
\email{chen@cs.ucr.edu}

\author{Franck Cappello}
\affiliation{%
  \institution{Argonne National Lab}
  \city{Lemont}\state{IL}
  \country{USA}
}
\email{cappello@mcs.anl.gov}

\renewcommand{\shortauthors}{Shixun Wu, Jinwen Pan, Jinyang Liu, Jiannan Tian, et al.}

\begin{abstract}
As high-performance computing architectures evolve, more scientific computing workflows are being deployed on advanced computing platforms such as GPUs. These workflows can produce raw data at extremely high throughputs, requiring urgent high-ratio and low-latency error-bounded data compression solutions. In this paper, we propose \textbf{{\thiswork}}, an optimized high-ratio GPU-based scientific error-bounded lossy compressor with a flexible, domain-irrelevant, and fully open-source framework design. Our novel contributions are: 1) We maximally optimize the parallelized interpolation-based data prediction scheme on GPUs, enabling the full functionalities of interpolation-based scientific data prediction that are adaptive to diverse data characteristics; 2) We thoroughly explore and investigate lossless data encoding techniques, then craft and incorporate the best-fit lossless encoding pipelines for maximizing the compression ratio of {\thiswork}; 3) We systematically evaluate {\thiswork} on benchmarking datasets together with representative baselines. Compared to existing state-of-the-art scientific lossy compressors, with comparative or better throughput than existing high-ratio scientific error-bounded lossy compressors on GPUs, {\thiswork} can achieve up to 249\% compression ratio improvement under the same error bound, and up to 215\% compression ratio improvement under the same decompression data PSNR.
\end{abstract}



\copyrightyear{2025}
\acmYear{2025}
\setcopyright{cc}
\setcctype{by}
\acmConference[SC '25]{The International Conference for High Performance Computing, Networking, Storage and Analysis}{November 16--21, 2025}{St Louis, MO, USA}
\acmBooktitle{The International Conference for High Performance Computing, Networking, Storage and Analysis (SC '25), November 16--21, 2025, St Louis, MO, USA}
\acmDOI{10.1145/3712285.3759798}
\acmISBN{979-8-4007-1466-5/2025/11}

\maketitle

\pagestyle{fancy}
\fancyfoot[C]{\fontsize{10pt}{12pt}\selectfont\thepage}  %
\renewcommand{\headrulewidth}{0pt}  %
\renewcommand{\footrulewidth}{0pt}  %

\newpage
\begingroup
\renewcommand{\contentsname}{\Large\textbf{Contents}}
\renewcommand{\addcontentsline}[3]{}%
\tableofcontents
\endgroup
\newpage

\section{Introduction}
\label{sec:intro}

There is no doubt that we are in the exascale computing era. Today, scientific applications tend to run on extremely large-scale environments, continuously producing massive amounts of scientific data.
The world's best exascale supercomputers, like Frontier~\cite{frontier}, Aurora~\cite{aurora}, and El Capitan~\cite{elcapitan}, which present peak performances around or over 1 Exaflops, have pushed the scientific computational power to a brand new magnitude, which is highly attributed to the cutting-edge GPU hardware that those supercomputers facilitate. Correspondingly, exascale scientific computing workflows have been established on GPU platforms, which include diverse scientific domains such as cosmology~\cite{nyx}, turbulence~\cite{gputurb}, and molecular dynamics~\cite{gpumd}.
However, higher computational power always leads to higher data generation speed, and the development of scientific computing workflows on GPU platforms is in great need of more effective data management solutions, which are critical to the storage, analysis, and transmission of exascale scientific data. For example, in~\cite{gputurb}, a GPU-accelerated turbulence simulation framework is proposed, which can generate data grids of 35 trillion grid points (128 TB in single-precision).
To this end, effective scientific data reduction techniques are required to be available on GPU platforms to alleviate the system bottleneck in I/O and storage with relatively low latency. The error-bounded lossy compression of scientific data can significantly reduce the data volume while preserving the per-data accuracy well, so it is commonly regarded as the best-fit data reduction solution for scientific data~\cite{sz3,zfp,cusz,sdrb,z-checker}.

Over the past years, various practices and deliverables of GPU-based scientific error-bounded lossy compression have emerged, projecting substantial potential for integrating them into GPU-based scientific computing workflows. For example, {\cusz}~\cite{cusz,cusz+,cuszi} and cuZFP~\cite{zfp,cuZFP} have been leveraged by existing scientific applications, and cuSZp~\cite{cuszp,huang2024cuszp2} is an excellent high-throughput data compressor to be leveraged in performance-favored GPU data compression. Nevertheless, a significant research gap remains in this field: \textbf{An open-source design for high-ratio scientific error-bounded lossy compression on GPUs is still missing}. Most existing works focus on high throughput, relying on low-cost and low-quality data processing algorithms. {\cusz}-I~\cite{cuszi}, as the first and the only existing GPU-based high-ratio-and-quality scientific data compressor, integrates an NVIDIA-proprietary lossless encoding module to achieve a compression ratio higher than 32, yet still exhibits a much lower compression ratio than CPU-based compressors like SZ3~\cite{sz3} and QoZ~\cite{sz3,HPEZ}. The most critical challenges of high-ratio scientific error-bounded lossy compression on GPU platforms are: 1) It is non-trivial to propose effective GPU-parallelization models and high-throughput implementations of high-ratio scientific data compression; 2) To optimize the GPU-based high-ratio scientific lossy compression pipelines, a thorough and in-depth benchmarking and analysis on the data decomposition and encoding modules on GPU platforms is required.

In this paper, committed to optimizing high-ratio scientific error-bounded lossy compression on GPU platforms, we propose {\thiswork}, which is a general, open-source, and highly effective solution. From the decomposition-encoding perspective of error-bounded lossy compression, based on the prior success of interpolation-based error-bounded lossy compressors~\cite{szinterp,qoz,HPEZ,cuszi}, we fine-craft a well-rounded GPU-parallelized interpolation-based data predictor, which is oriented toward maximizing the lossless compressibility of integer decomposition data. Furthermore, from systematic investigation, evaluation, and benchmarking on GPU-based numerical lossless encoders, we establish the best lossless pipelines for {\thiswork}, which fully leverages different aspects of redundancy information in the data, ensuring the high-magnitude compression ratio among all use cases. Our contributions are as follows:

\begin{itemize}
\item In {\thiswork}, we further optimize the high-throughput interpolation-based data prediction on GPU platforms, proposing a new scientific data prediction module with multiple interpolation schemes and auto-tuned configurations;
\item After an in-depth investigation of existing numerical lossless encoders, we craft and incorporate the best-fit lossless encoding pipelines for maximizing the compression ratio of {\thiswork}.
\item We perform a comprehensive evaluation of {\thiswork} on 6 real-world scientific datasets. {\thiswork} exhibits an outperforming compression ratio and quality compared to all existing state-of-the-art. {\thiswork} can achieve up to 249\% compression ratio improvement over existing scientific lossy compressors under the same error bound and up to 215\% compression ratio improvement under the same decompression data PSNR.
\end{itemize}

We organize the rest of this paper as follows: \S\ref{sec:related} discusses the related work of this paper. \S\ref{sec:background} formulates our research target and the background of interpolation-based scientific error-bounded lossy compression on GPUs. In \S\ref{sec:framework}, we propose the framework design and module composition of {\thiswork}. \S\ref{sec:details} demonstrates our detailed designs for optimizing {\thiswork}. In \S\ref{sec:evaluation}, we present the evaluations for {\thiswork}. \S\ref{sec:conclusion} concludes this work and discusses future plans.

\section{Related Work}
\label{sec:related}

\subsection{Scientific error-bounded lossy compression}

Error-bounded lossy compression has been regarded as the best-fit strategy for scientific data size reduction. Existing error-bounded lossy compressors have proposed diverse technical solutions, and present different capabilities on use cases and metrics. Among the most representative works, SZ-family~\cite{sz-auto,sz3,qoz,HPEZ} follows a hybrid design of spline interpolation \cite{szinterp}, Linear Regression, and Lorenzo extrapolation~\cite{sz17}. ZFP~\cite{zfp} and SPERR~\cite{SPERR} process data with transforms and specialized encoding schemes. TTHRESH~\cite{ballester2019tthresh} leverages high-order singular value decomposition to effectively decorrelate the input data. To maximally reduce the compressed data size, there are various deep-learning-integrated frameworks for scientific error-bounded lossy compression~\cite{ae-sz,SRNN-SZ,huang2022compressing,lu2021compressive,hayne2021using,liu2024enhancing,li2024attention}. To optimize the compression throughput, compressors like SZx~\cite{szx} and SZp (FZ-light)~\cite{hzccl} reduce their lossy data processing scheme to very-simplified offset-computation and bit-analysis.

\subsection{Error-bounded lossy compression on GPUs}

Based on the aforementioned scientific error-bounded lossy compression frameworks, several GPU-based scientific lossy compressors have been proposed for several purposes: (1) They can deliver much higher magnitudes of data throughput than CPU-based compressors; (2) They can be directly integrated into existing and emerging scientific computing applications on GPU platforms with little system latency and overhead. Several examples are:

\begin{itemize}
  \item \textbf{{\cusz}:} {\cusz}~\cite{cusz,cusz+,cuszi} was proposed in 2020, revised in 2021, and updated in 2024. The latest version of {\cusz} merges the recently proposed {\cusz}-I~\cite{cuszi} and the original Lorenzo-based {\cusz}~\cite{cusz,cusz+}. It integrates 2 data predictors (Lorenzo predictor and interpolation predictor) and then applies Huffman encoding (and optionally, NVIDIA Bitcomp) as the lossless module.
  \item \textbf{cuSZp2:} cuSZp2~\cite{huang2024cuszp2} is an end-to-end performance optimized scientific error-bounded lossy compressor on GPU platforms. It leverages 1D offset prediction and fixed-length encoding for high-speed scientific data compression.
  \item \textbf{cuZFP:} cuZFP~\cite{zfp,cuZFP} performs scientific error-bounded lossy compression with discrete orthogonal transform and embedding encoding. It has a balanced compression ratio and throughput.
  \item \textbf{FZGPU:} Derived from {\cusz}, FZGPU alters the compression pipeline by replacing the lossless encoding stage with bit-shuffle and dictionary encoding, aiming to deliver higher throughput.
  \item \textbf{Others:} cuSZx~\cite{szx} features a monolithic design, delivering extremely high throughput at the cost of lower data quality and compression ratio. MGARD-GPU~\cite{mgard-latest} is another GPU compressor practice based on the MGARD compression algorithm~\cite{MGARD}.
\end{itemize}
\subsection[Lossless encoding in scientific context]{Lossless encoding in scientific lossy compression}

In the past years, lossless compression (encoding) techniques have not only served as standalone data compressors but have also been widely adopted by scientific error-bounded lossy compressors to improve data reduction rates~\cite{liu2021improving}. From gzip~\cite{gzip} in SZ1.0, Zstd~\cite{zstd} in SZ3, SPECK~\cite{speck} in SPERR, to NVIDIA Bitcomp in {\cusz}-I~\cite{cuszi}, diverse lossless encoders have enhanced the scientific error-bounded lossy compression. Among those encoders, there are entropy-based encoders that compress the data by token frequency information (such as Huffman encoding), redundancy-based encoders that compress the data by reducing repeat patterns (such as run-length encoding), and a mixture of those 2 techniques (such as Zstd, Deflate~\cite{deutsch1996deflate}, etc.). On GPU platforms, there are also multiple open-source and proprietary lossless encoders for scientific numerical data~\cite{chen2024fcbench}, including but not limited to GPULZ~\cite{zhang2023gpulz}, ndzip~\cite{ndzip-gpu}, and nvCOMP~\cite{nvcomp}. The recently developed LC framework~\cite{LC} integrates over 10 long-standing components (e.g., mutators, shufflers, reducers) used in constructing lossless compressors, supporting various symbol widths. This framework enables users to traverse diverse component combinations for the files requiring compression and customize compressors with an arbitrary number of stages at a fine-grained level while supporting heterogeneous compression/decompression across GPUs and CPUs.

\section{Research Background}
\label{sec:background}

\subsection[Problem formulation]{Problem formulation of scientific error-bounded lossy compression}

We mathematically formulate our research problem: scientific error-bounded lossy compression on GPU platforms. Generally, given an input scientific data $X = \{x_i\}$, a pair of compressor and decompressor $(\operatorname{Cmp}, \operatorname{Dec})$ converts $X$ to compressed data $Z = \operatorname{Cmp}(X)$ and decompressed data $X^{\prime} = \{x_i^{\prime}\} = \operatorname{Dec}(Z)$. The optimization of the compression ratio, which is defined by the input data size divided by the compressed data size, is described as follows:
\begin{equation}
  \label{eq:general}
  \begin{split}
    &\text{maximize\ } \frac{|X|}{|Z|} \\
    \text{s.t.\ } & Z = \operatorname{Cmp}(X),X^{\prime} = \operatorname{Dec}(Z) \\
    \text{and \ } & \| X-X^{\prime} \|_{\infty} = \max |x_i-x_i^{\prime}|\leq \epsilon \\
  \end{split}
\end{equation}
In Eq.~\ref{eq:general}, the upper bound of point-wise data error $\epsilon$ is named the error bound in the compression task. We further specify this problem in 2 aspects: 1) Regarding the GPU-based error-bounded lossy compression, the whole compression pipeline should be highly parallelized on the hardware, presenting high computing throughputs (often ten or hundreds of Gigabytes of data per second in modern platforms); 2) We observe that scientific error-bounded lossy compressors mostly follow a framework design of lossy decomposition plus lossless encoding (detailed in Section~\ref{sec:framework}). The input floating-point data is first lossily decomposed into integer data with high compressibility, and the integers are then further losslessly encoded to achieve size reduction. Eq.~\ref{eq:specific} involves this framework scheme, formulating our research target. We would like to jointly optimize the integer decomposition process (Dcp) and the lossless encoding (Enc) process while preserving the required computing throughput ($\operatorname{Speed}(\operatorname{Dcp}, \operatorname{Enc})$) with speed higher than a threshold $S_t$ (we initially set it to a comparable speed to {\cusz}-I):
\begin{equation}
\label{eq:specific}
\begin{split}
&\operatorname{Dcp},\ \operatorname{Enc} = \arg \max_{\operatorname{Dcp},\ \operatorname{Enc} }  \frac{|X|}{|\operatorname{Enc}(\operatorname{Dcp}{X})|} \\
s.t.  \   &\left\|X-X^{\prime}\right\|_{\infty} = \max |x_i-x_i^{\prime}|\leq \epsilon \\
and \ & \operatorname{Speed}(\operatorname{Dcp}, \operatorname{Enc}) \geq S_t
\end{split}
\end{equation}

\subsection[Interpolation-based compression on GPUs]{Interpolation-based error-bounded lossy compression on GPUs}

\begin{figure}
\centering
\includegraphics[width=.9\linewidth]{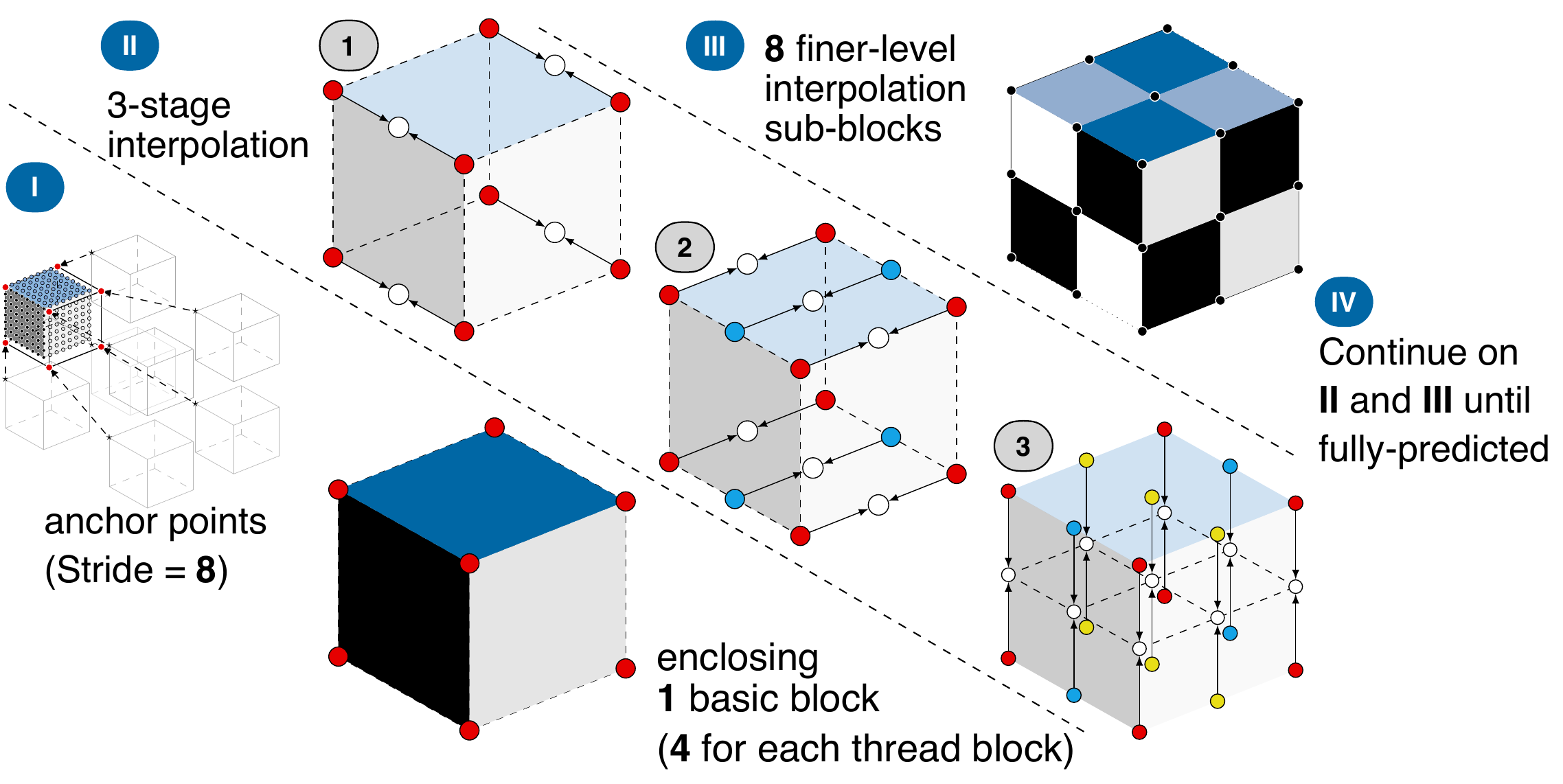}
\caption{Design overview of {\cusz}-I (interpolation-based scientific error-bounded lossy compression on GPUs).}
\label{cuszhi::fig::interp}
\end{figure}

In this subsection, we introduce the interpolation-based error-bounded lossy compression on GPUs (specifically {\cusz}-I \cite{cuszi}) as our research background, as {\thiswork} also integrates an interpolation-based data predictor. The core design of {\cusz}-I is illustrated in Figure~\ref{cuszhi::fig::interp}. To ensure parallelization, {\cusz}-I partitions the input data into small chunks (e.g., $33 \times 9 \times 9$ for 3D data), each composed of 4 basic blocks (shown in Figure~\ref{cuszhi::fig::interp}), and allocates each chunk to a thread block. Leveraging spline interpolations, a thread block predicts the allocated data chunk from a losslessly stored sparse anchor grid (in the stride of 8 on each dimension) and then quantizes the prediction errors to integers for the subsequent lossless size reduction process. The interpolations follow a hierarchical scheme, performing interpolation in different sequential steps, from high levels (with large interpolation strides) to low levels (with small interpolation strides) along all dimensions. The computations are fully parallelized on the threads within each interpolation level and dimension. After acquiring the quantized integer errors (also known as the quantization codes), {\cusz}-I applies Huffman encoding and an optional proprietary NVIDIA Bitcomp~\cite{nvcomp} lossless compression module to finalize the size reduction.

{\cusz}-I (especially its Bitcomp-integrated version) has achieved state-of-the-art compression ratio and data quality in GPU-based scientific error-bounded lossy compression. In this paper, we will keep focusing our compressor design on interpolation and integer encoding techniques due to the unique potential of interpolation in proposing high-accuracy data predictions with relatively low computational costs. Furthermore, we will address several existing critical limitations of interpolation-based scientific data compression on GPUs, including but not limited to 1) suboptimal data-partition and anchor-sampling strategy; 2) outdated design of the interpolation process and its auto-optimization; 3) lack of an effective and open-source lossless encoding solution for quantized errors.

\section{The Framework Design of \thiswork}
\label{sec:framework}
\begin{figure}
  \centering
  \includegraphics[width=0.99\linewidth]{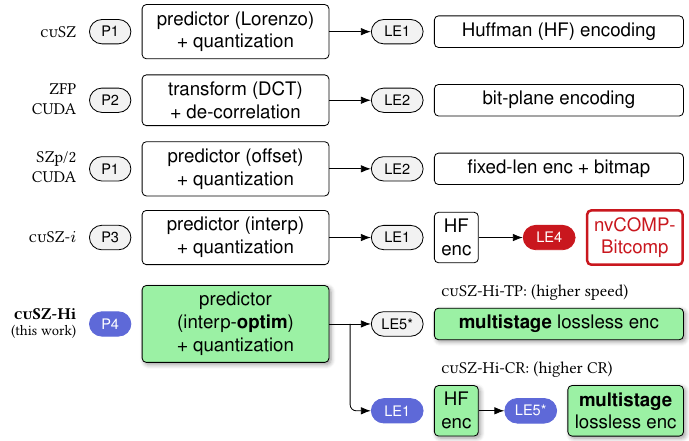}
  \caption{Decomposition-encoding-based framework design of {\thiswork}, compared to existing works.}
  \label{cuszhi::fig::pipelines}
\end{figure}

The compression framework design of {\thiswork} is proposed in Figure~\ref{cuszhi::fig::pipelines}, comparing it to other existing representative GPU-based scientific error-bounded lossy compressors. Although those compressors have diverse compression framework designs, they all share a uniform scheme, which is lossy data decomposition and lossless integer encoding, as shown in Figure~\ref{cuszhi::fig::pipelines}. The first module of each framework decorrelates the input floating-point data, producing integers as intermediate results. Those integers have higher compressibility than the original input, contain reduced information but preserve the error bound, and are generated by different data processing techniques such as direct value quantization (in cuSZp2), Lorenzo extrapolation (in {\cusz}), and spline interpolation (in {\cusz}-I). Afterward, the lossless encoding modules in each compressor reduce the size of the intermediate integers without any information loss, ensuring that the final output serves as the error-bounded compression result. We detail the framework modules of {\thiswork} as follows and explain why they bring an optimized design regarding compression ratio and quality.

\begin{itemize}
  \item \textbf{Lossy data decomposition:} {\thiswork} adopts spline-interpolation-based data prediction, which features an advanced design over the predictor in {\cusz}-I~\cite{cuszi}. Spline-interpolation-based data prediction has been proven to be the best-fit data decomposition technique for scientific error-bounded lossy compression because it can achieve both high data prediction accuracy and computation speed~\cite{szinterp,qoz,HPEZ,cuszi}. In contrast, compressors with other lossy decomposition techniques, such as {\cusz}, cuSZp2, and cuZFP, suffer from a significantly lower compression ratio than {\cusz}-I.
    Based on the high-throughput parallelized implementation of spline interpolations proposed by~\cite{cuszi}, {\thiswork} further tunes and optimizes the GPU-based spline interpolation design to maximize the compression ratio acquired.
  \item \textbf{Lossless integer encoding:} To maximize the compression ratio, a well-rounded lossless encoding pipeline is necessary, which should be able to leverage the data correlation information on both the token-level and bit-level.
    Unfortunately, among existing compressors, {\cusz} only integrates the frequency-based Huffman encoding, and cuSZp2/cuZFP merely applies fixed-length bit-wise de-redundancy encodings. {\cusz}-I managed to combine both Huffman encoding and de-redundancy encoding, but the NVIDIA Bitcomp module it leverages is a proprietary product with limited usability. In {\thiswork}, after synthetically investigating and benchmarking available lossless encoding solutions, {\thiswork} proposes new open-source lossless pipelines that jointly facilitate Huffman encoding and high-performance multi-stage lossless modules. Moreover, as indicated in Figure~\ref{cuszhi::fig::pipelines}, {\thiswork} integrates 2 different lossless pipelines, which can be flexibly selected according to the user's requirements, presenting a higher compression ratio (in \textbf{{\thiswork}-CR} mode) or throughput (in \textbf{{\thiswork}-TP} mode).
\end{itemize}

\section{Synergistic Lossy-Lossless Module Design}
\label{sec:details}
As Figure~\ref{cuszhi::fig::pipelines} indicates, optimizing the floating data compression ratio of scientific error-bounded lossy compression pipelines is equivalent to improving the compression for the `` lossy decomposed" integer array (in {\thiswork}, it is the array of quantized predictor errors). To this end, there are two different but synergistic solutions: Increasing the compressibility of the integer array and promoting the capability of lossless integer encoding modules. In this section, we will detail our new GPU-based compression module designs regarding those 2 methodologies.

\subsection[Improving interpolation-based prediction]{Improving interpolation-based lossy data prediction}
\label{sec:lossy}
Because lossless data encoding algorithms mainly leverage 2 types of data information: token-wise entropy and bit-wise redundancy, there are also 2 direct solutions for improving the compressibility of quantization codes from the {\thiswork} interpolation-based data predictor. First, we present improved data prediction accuracy and anchor-point rate, resulting in a more concentrated distribution of quantization codes around zero, which delivers lower entropy. Second, based on the characteristics of interpolation-based data predictors, we reorder the quantization code array using a fixed mapping, grouping similar values to enhance their compressibility.
\subsubsection{Reorganized data and anchor partitions}
\label{sec:partition}
\begin{figure}
\centering
\includegraphics[width=0.99\linewidth]{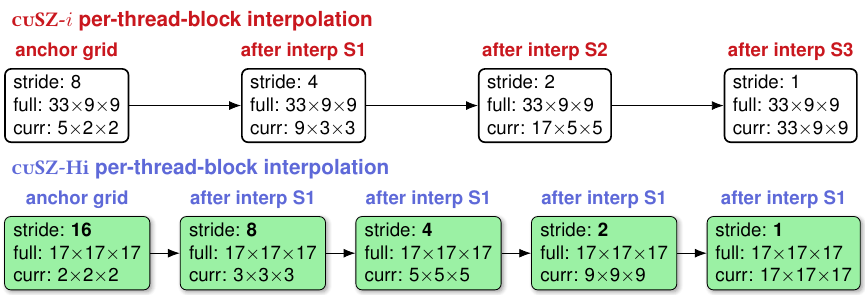}
\caption{Per-thread-block anchor and interpolation partition in {\cusz}-I and {\thiswork} on 3D data.}
\label{cuszhi::fig::partition}
\end{figure}

In Figure~\ref{cuszhi::fig::partition}, we present the details of {\thiswork} thread-block-wise data partition, anchor layout, and interpolation workloads in a 3D example, comparing it to {\cusz}-I. Specifically: 1) In {\thiswork}, the 3D partitioned data block to be processed by each thread block has a size of $17^3$, including all anchors) instead of $33 \times 9 \times 9$ for {\cusz}-I. 2) On each dimension, the stride of losslessly stored anchor points is 16 in {\thiswork} instead of $8$ in {\cusz}-I. Due to this, the interpolations for 3D data blocks in {\thiswork} follow a 4-level hierarchy instead of the {\cusz}-I 3-level interpolation. We justify our new design as follows: First, a dimensionally isotropic data partition can be better adaptive to different interpolation computation orders, which will be detailed in \S\ref{sec:att}. Second, the data block size in {\thiswork} is double the original data block size in {\cusz}-I, which better exploits the shared memory and cache in GPU architectures and brings more available data points in the prediction process for improving the data prediction accuracy~\cite{HPEZ}. Moreover, in our preliminary experiments, further enlarging the data block yields limited improvement in compression ratio, while significantly reducing the compression speed. Last, according to preliminary experiments, anchor points in {\cusz}-I occupy a considerable portion of the compression data size. Reducing the number of anchor points (to $\frac{1}{8}$ of {\cusz}-I) can effectively reduce the anchor storage overhead while preserving data prediction quality.

\subsubsection{Parallelized multi-dimensional interpolation}
\begin{figure}[ht]
\centering
\hspace{-5mm}
\subfigure[\textbf{1D interpolation}]
{
\raisebox{-1cm}{\includegraphics[scale=0.36]{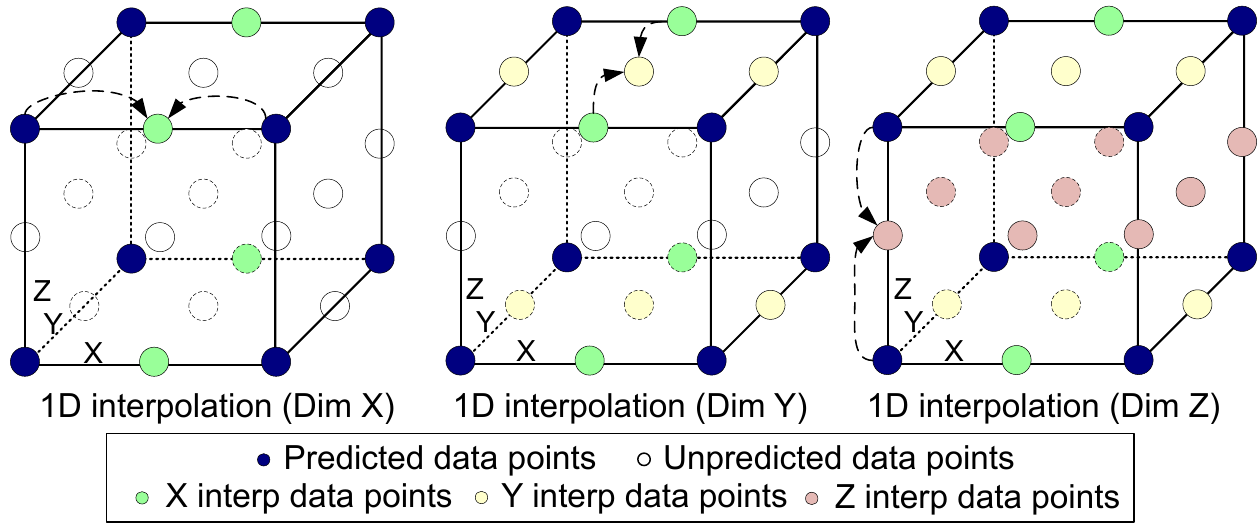}}%
}
\hspace{-5mm}

\hspace{-5mm}
\subfigure[\textbf{Multi-dimensional interpolation}]
{
\raisebox{-1cm}{\includegraphics[scale=0.36]{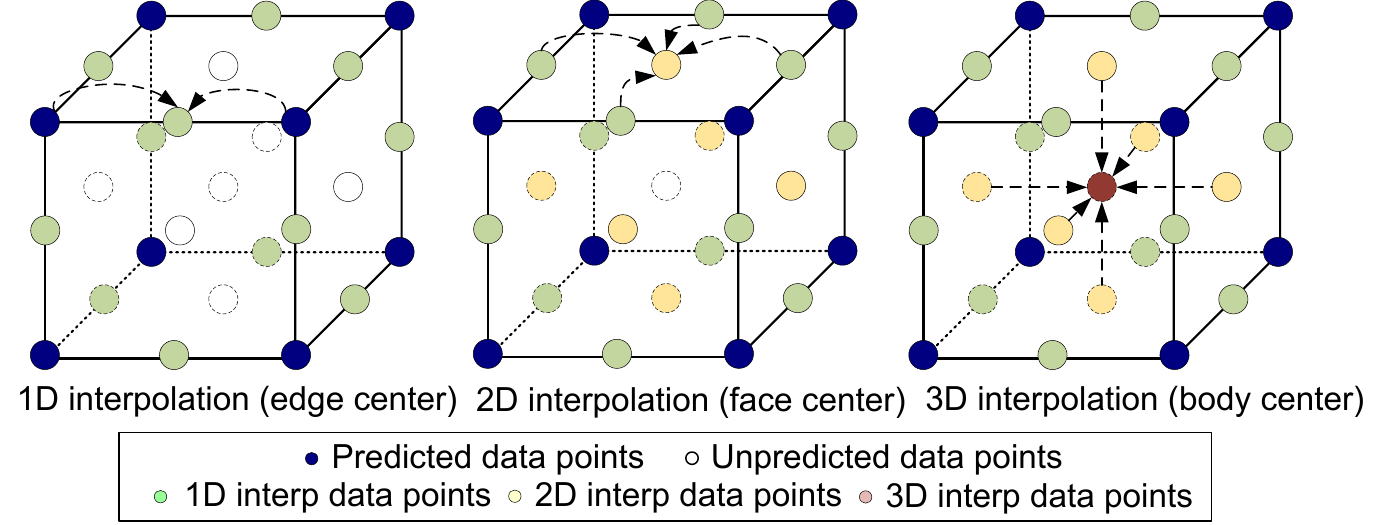}}%
}
\hspace{-5mm}
\vspace{-1mm}
\caption{GPU-based multi-dimensional interpolation (v.s. 1D interpolation). Interpolations for data points in the same colors are fully parallelized.}
\label{cuszhi::fig::mdi}
\end{figure}

Regarding the GPU-parallelized design of interpolation-based data prediction, {\thiswork} also introduces a new scheme, which is featured in Figure~\ref{cuszhi::fig::mdi}.  To further leverage the data correlation along multiple data dimensions, adopting the design concepts proposed in~\cite{HPEZ}, {\thiswork} crafts a new parallelized, multi-dimensional-interpolation-based prediction scheme in each partitioned data block (Figure~\ref{cuszhi::fig::mdi}~(b)). Unlike original 1D interpolation dimension sequences ((Figure~\ref{cuszhi::fig::mdi}~(a)), on each level, parallelized multi-dimensional interpolations follow a process with an increasing number of dimensions used in the interpolation (1D $\to$ 2D $\to$ 3D). This scheme is isotropic along all dimensions (free of dimension sequence selection) and better leverages the multidimensional data correlation than the existing one-dimensional spline interpolation scheme. When predicting a specific data point along multiple dimensions, the prediction value is the average of the interpolation predictions along those dimensions. Moreover, to improve the prediction quality, only prediction values with the highest spline order will be used and averaged in the prediction. For example, two cubic spline interpolations along two dimensions in {\thiswork} will be averaged as the final data prediction, but if one is cubic and the other is linear/quadratic, {\thiswork} will only use the cubic spline interpolation as the prediction result.

\subsubsection{Workload-balanced auto-tuning}
{\thiswork} introduces multiple interpolation splines and schemes, which adapt differently to the characteristics of the input data. To this end, an online dynamic selection of those splines and schemes will be critical in optimizing the compression ratio of {\thiswork} on diverse data inputs. Previously, {\cusz}-I merely conducted fast profiling on sampled input data points to determine the best-fit interpolation spline and scheme, which is insufficient to make an accurate selection for interpolation configuration. Following the design proposed in \cite{qoz,HPEZ}, the compression pipeline of {\thiswork} includes a full-functional module of interpolation auto-tuning, which is still lightweight enough, with well-balanced workloads on thread blocks to minimize the overhead in GPU-parallelized compression computing. Specifically, {\thiswork} first uniformly samples data blocks (with the same size as per-thread-block-data), with a total volume of $0.2\%$ of the whole data. In the auto-tuning kernel, each sampled data block is distributed to multiple thread blocks to perform interpolations at different levels using various splines and schemes.

To optimize the performance of the auto-tuning kernel, balancing the number of thread blocks used and the per-block workload, {\thiswork} manages the interpolation test differently at different levels. For level 4 and level 3 interpolations, which have limited computational cost, interpolation tests with all possible configurations (spline $+$ scheme) are performed on the same threadblock. For level 2 interpolation, the tests are equally allocated to 2 thread blocks. For level 1 interpolation, which requires intensive computation, {\thiswork} leverages 6 thread blocks to test one configuration on each. %
After performing the interpolation tests on each level and with each configuration, {\thiswork} collects the aggregated data prediction errors from each test and selects the interpolation configuration separately on each level that has the minimized test error.

\subsubsection{Mapping-based quantization codes reordering}
The Huffman encoding module (which is adopted by SZ3, {\cusz}, and {\cusz}-I) is entropy-based, which means that its compression ratios on lossy-predicted quantization codes are solely dependent on the frequency distribution of the tokens (i.e., values of the codes). However, it cannot exploit the sequential order of the tokens and the bit-wise redundancies in the tokens. To this end, {\thiswork} incorporates additional bitwise-redundancy-based lossless encoding modules (detailed in \S\ref{sec:lossless}), and the quantization codes are also reversibly reordered to optimize the de-redundancy encoding.

As previously mentioned, in {\thiswork}, the data predictions are performed by spline interpolations with diverse strides. For example, data points whose coordinates have at least one odd value are predicted by interpolation with a stride of 1, and the rest of the data are predicted by interpolation with larger strides (2 to 8). In existing {\cusz}-I, the quantized prediction errors are gathered in an array that has the same layout as the input data, and the coordinates of each element remain the same as the corresponding input data point. Therefore, when the quantization code array is flattened to a 1D sequence for lossless encoding, the quantization codes from different interpolation levels and strides will get shuffled in the sequence. However, several existing research works \cite{qoz,HPEZ,cuszi} have observed and proved that the prediction accuracy of an interpolation-based data predictor is highly relevant to the interpolation stride, for which larger interpolation strides lead to higher data prediction errors. Therefore, quantization codes from large-stride interpolations often present larger absolute values and exhibit different distributions from quantization codes generated by small-stride interpolations. Those insights into the distribution of the quantization codes drive us to rearrange the to-be-encoded quantization code sequence by separating quantization codes from different interpolation strides.

\begin{figure}[ht]
\centering
\raisebox{-1mm}{\includegraphics[scale=0.33]{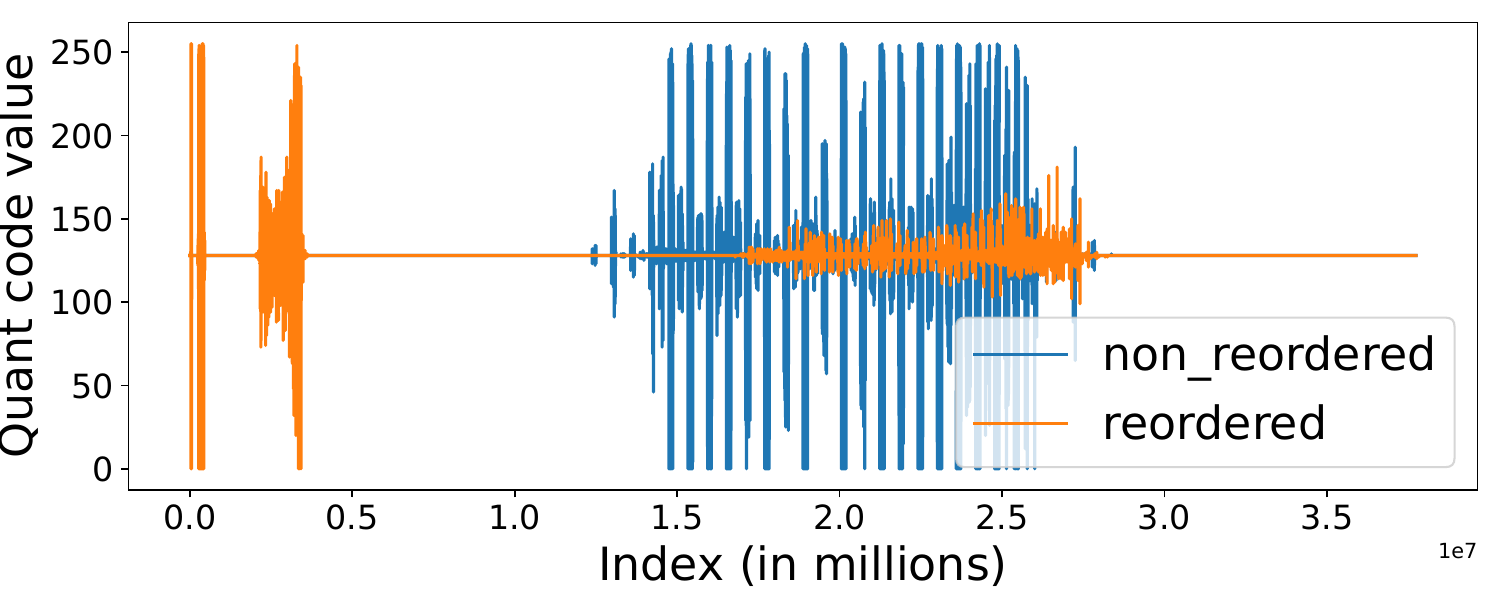}}
\caption{Comparison between original and reordered quantization code sequences (from compressing Miranda-Pressure data snapshot with error bound $\epsilon$ = 1e-3).}
\label{fig:reorder}
\end{figure}

Specifically, after acquiring the multi-dimensional quantization code array (the coordinates are the same as the data array), {\thiswork} maps the array to a reordered 1D sequence for lossless encoding, in which quantization codes for each interpolation level are grouped in adjacent indices. $A$ is the stride of anchor points, like 16, $(d_x,d_y,d_z)$ are the global data dimension sizes. For a quantization code with coordinate $(x,y,z)$ in the 3D code array, its index $I(x,y,z)$ in the mapped 1D sequence will be:
{\footnotesize
\begin{equation}\label{eq:reorder}
\begin{split}
  {I}(x,y,z) =
  & \operatorname{prefix}(\ell)+xd_yd_z+yd_z+z
  - \left\lfloor\frac{x+1}{2}\right\rfloor\left\lfloor\frac{d_y+1}{2}\right\rfloor\left\lfloor\frac{d_z+1}{2}\right\rfloor
  \\
  & - \mathbf{1}_{\{x\,\text{even}\}}\left\lfloor\frac{y+1}{2}\right\rfloor\left\lfloor\frac{d_z+1}{2}\right\rfloor
  - \mathbf{1}_{\{x,y\,\text{even}\}}\left\lfloor\frac{z+1}{2}\right\rfloor
  \\
  \text{interp level \ }\ell =
  & \max \left\{ \ell \in \mathbb{N},\ 0 \leq \ell \leq \log_2{A},\ 2^\ell \ |\ x,y,z\right\},
  \\
  \operatorname{prefix}(\ell) =
  &
  \begin{cases}
    f_x(\ell) f_y(\ell) f_z(\ell), & \ell<\log_2{A}  \\
    0, & \ell = \log_2{A}
  \end{cases} \\
\end{split}
\end{equation}%
with
\begin{equation}
\begin{aligned}
  f_x(0) = \left\lfloor\frac{d_x+1}{2}\right\rfloor, &\ f_x(k+1) = \left\lfloor\frac{f_x(k)+1}{2}\right\rfloor,  \\
  f_y(0) = \left\lfloor\frac{d_y+1}{2}\right\rfloor, &\ f_y(k+1) = \left\lfloor\frac{f_y(k)+1}{2}\right\rfloor,  \\
  f_z(0) = \left\lfloor\frac{d_z+1}{2}\right\rfloor, &\ f_z(k+1) = \left\lfloor\frac{f_z(k)+1}{2}\right\rfloor.
\end{aligned}
\end{equation}%
}%
In Eq.~\ref{eq:reorder}, $\ell$ indicates the interpolation level for predicting the data point with coordinate $(x,y,z)$, and $\operatorname{prefix}(\ell)$ is the number of data points on higher interpolation levels. According to Eq.~\ref{eq:reorder}, quantization codes from the larger interpolation strides will appear first in the 1D code sequence (with smaller sequence index $I(x,y,z)$), followed by codes from smaller interpolation strides. In Figure~\ref{fig:reorder}, we present the distribution of quantization code values by indices in the code sequences, both from direct flattening and {\thiswork} reordering. In the non-reordered code sequence, the code values oscillate intensively over a wide range. Quantization code reordering transfers most outlier values to a short range (at the beginning of the sequence), thereby significantly increasing the overall smoothness of the sequence and improving the compression ratio.

\label{sec:att}
\subsection{Optimizing lossless ratio and throughput}
\label{sec:lossless}

Past research over years~\cite{liu2021improving,cusz+} and recently proposed~\cite{chen2024fcbench,LC} have proved that the evolving lossless data encoding techniques have great potential in improving the scientific lossy compression over various aspects, including both compression ratio and throughput. According to existing work~\cite{cuszi} and our investigation, one key finding is that \textbf{Multiple existing scientific lossy compressors have not fully utilized the correlation information in the original data, leaving undealt compressibilities in the output data.} In Table~\ref{tab:bitcompcr}, we present a showcase, listing the lossless compression ratio by NVIDIA Bitcomp~\cite{nvcomp} of several compressed outputs from scientific error-bounded lossy compressors, including baselines (detailed in \S\ref{sec:related} and \S\ref{sec:setup}) and our proposed {\thiswork}. Most of the compression outputs can still be highly compressed by NVIDIA Bitcomp, reducing their sizes to $10\% \sim 40\%$ of the original size, which is the reason why {\cusz}-I directly integrated NVIDIA Bitcomp as a module of its pipeline. Unfortunately, Bitcomp is NVIDIA-proprietary, so integrating it into the compression framework will bring severe restrictions on its extension, customization, and (cross-platform) deployment. To maximize the scientific lossy compression ratio on GPUs, achieving minimal information redundancy while preserving the usability of the whole compression framework, as {\thiswork} does, we need a new solution for the lossless pipeline, which will be detailed step-by-step in the next.

\begin{table}[ht]
\FloatBodyStyle
\caption{The NVIDIA Bitcomp compression ratio on the compressed data from different error-bounded lossy compressors (Nyx~\cite{nyx} dataset, error bound = 1e-2).}
\resizebox{\linewidth}{!}{
\begin{tabular}{| l@{}c | lc |}
  \hline
  \multirow{2}{*}{\textbf{compressor}}  & \footnotesize \textbf{Bitcomp CR on}&
  \multirow{2}{*}{\textbf{compressor}}  & \footnotesize \textbf{Bitcomp CR on}\\[-1.ex]
  & \textbf{comp'ed data} &
  & \textbf{comp'ed data} \\ \hline\hline
  {\thiswork}-CR   & 1.03  &   {\cusz}-L      & 2.37 \\ \hline
  {\thiswork}-TP   & 1.06  &   cuSZp2      & 3.33 \\ \hline
  {\cusz}-I (w/o Bitcomp)     & 9.62  &   FZ-GPU      & 3.33 \\ \hline
\end{tabular}}
\label{tab:bitcompcr}
\end{table}

\subsubsection{Minimizing quantization code length}

The first important design detail of {\thiswork} lossless pipeline is that, in the integer quantization code sequence, each value has been resized to the minimized one-byte-width, processed in the uint8\_t format (the out-of-range outliers will be collected and stored separately). Unlike other scientific lossy compressors that quantize data offsets~\cite{cuszp,huang2024cuszp2} or introduce inaccurate extrapolation errors~\cite{cusz}, interpolation-based data compressors typically produce quantization codes with a highly concentrated distribution, including very few large values~\cite{cuszi}. Consequently, applying a short data format to store the quantization codes will accelerate the encoding speed and simplify bit patterns without high overheads for recording outliers.

\subsubsection{Investigation and benchmarking of lossless encoders}

\newcommand{\LLStage}[1]{\texttt{#1}}

To figure out the best-fit lossless pipeline for encoding the {\thiswork} quantization codes, we conduct a systematic benchmarking of compressing them with a large variety of GPU-based numerical data encoders introduced in existing works~\cite{chen2024fcbench,LC}. The LC framework~\cite{LC} integrates a large variety of lossless encoding algorithms and modules. It can search for and combine existing transforming and reducing components to form synthetic lossless pipelines with a customizable number of stages. NVIDIA nvCOMP~\cite{nvcomp} includes a series of NVIDIA-proprietary GPU-optimized implementations of lossless compression algorithms. GPULZ~\cite{zhang2023gpulz} is an open-source implementation of the LZSS algorithm for GPUs. ndzip~\cite{ndzip-gpu} is an open-source high-throughput lossless compressor that supports multi-dimensional data and is compatible with both GPU and CPU architectures.

\begin{figure}
\includegraphics[width=\linewidth]{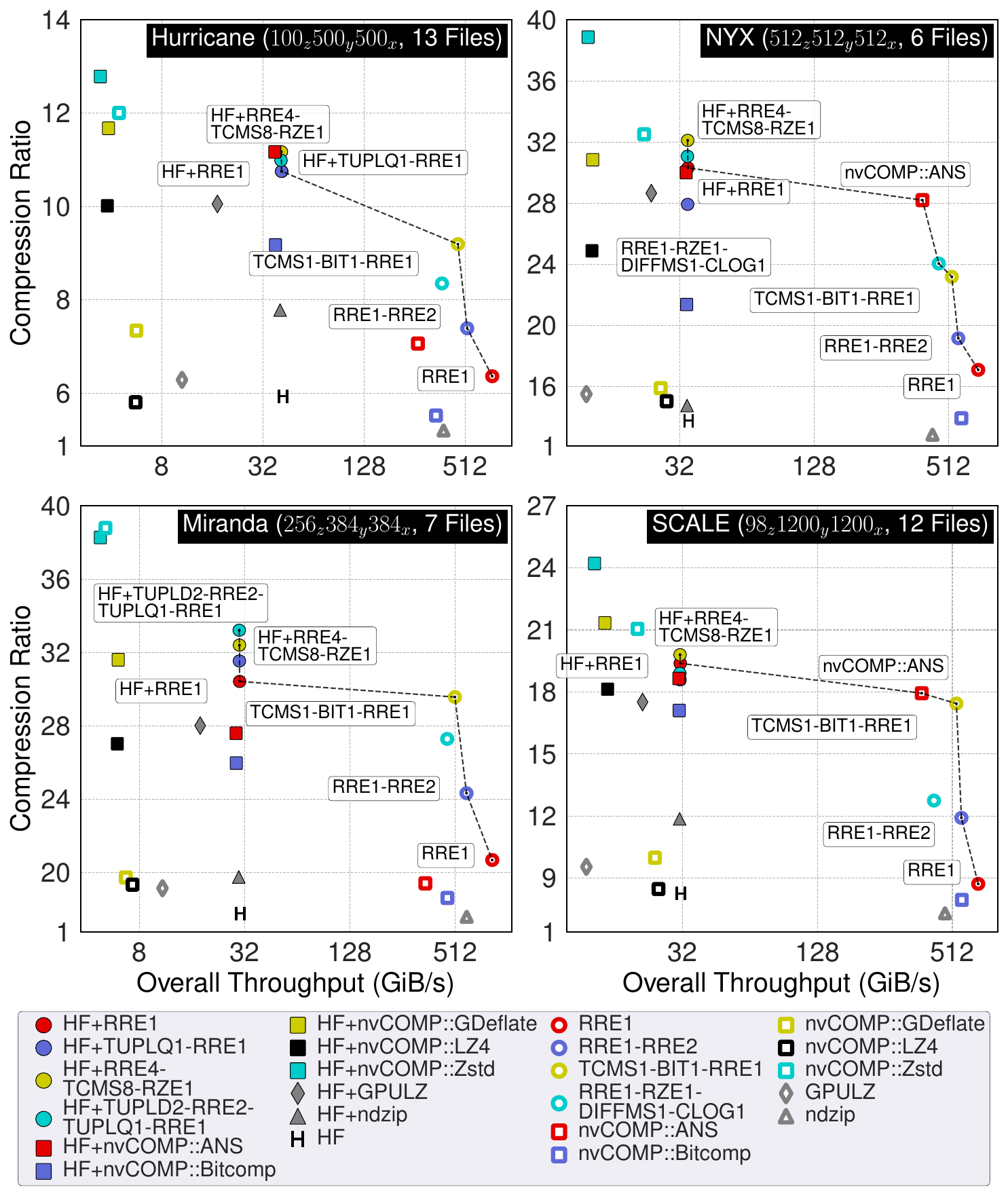}
\caption{Benchmarking of lossless compressors with quantization codes  (error bound = 1e-3) as input on RTX 6000 Ada. The dashed lines represent Pareto frontiers, excluding points with an overall throughput of less than 25 GiB/s.}
\label{cuszhi::fig::lossless_1e-3}
\end{figure}

In our benchmarking, the evaluated lossless pipelines are collected as follows: First, incorporating Huffman encoding as a "preprocessor" in the lossless compression pipeline has been a widely adopted scheme in existing compressors~\cite{sz3,HPEZ,cuszi}, so in addition to the standalone lossless modules, we also include a set of Huffman-incorporated variants corresponding to each of them. Second, for the LC framework, we perform preliminary experiments on several datasets and select 8 representative and adaptive pipelines (\LLStage{RRE1}, \LLStage{RRE1-RRE2}, etc., as shown in Figure~\ref{cuszhi::fig::lossless_1e-3}) with 1/2/3/4 stages (as indicated by Figure~\ref{cuszhi::fig::lossless_1e-3}, pipelines with more stages are not necessary).
The benchmarking is conducted on 4 scientific datasets~\cite{sdrb} on a computing platform facilitated with NVIDIA RTX 6000 Ada GPUs.

The compression ratio and overall (compression-decompression) throughput of each lossless pipeline are shown in Figure~\ref{cuszhi::fig::lossless_1e-3}, with Pareto frontiers marked (solutions with throughput less than 25 GiB/s are not included in the frontiers because they bring unacceptable time overheads to the full compression pipeline).
Regarding open-source solutions with acceptable compression throughput, we find that the combination of \textbf{Huffman encoding} (hereinafter \LLStage{HF}) and the \LLStage{\bfseries RRE4-TCMS8-RZE1} pipeline achieves high compression ratios on {\thiswork} quantization codes.
The Zstd~\cite{zstd} compressor in NVIDIA nvCOMP achieves the highest compression ratio. However, not only is this implementation NVIDIA-proprietary, but it also consists of multiple self-contained encoders with coarse component configuration granularity. Consequently, its throughput is significantly low in most scenarios, making it impractical for {\thiswork}. The nvCOMP::ANS/Bitcomp and GPULZ may present decent compression ratios in some scenarios, but they show no advantages over the multi-stage LC pipelines. The GDeflate/LZ4/ndzip/Huffman-only lossless solutions exhibit very poor compression ratios and/or throughputs, being infeasible for {\thiswork} usage.

According to the benchmarking results and analysis, we adopt the \LLStage{\bfseries HF-RRE4-TCMS8-RZE1} lossless pipeline in {\thiswork}, forming its compression-ratio-preferred mode, \textbf{{\thiswork}-CR}. However, Huffman encoding on GPUs also brings a performance bottleneck in the compression framework and lowers the overall compression throughput~\cite{9820677}. To this end, regarding Figure~\ref{cuszhi::fig::lossless_1e-3}, {\thiswork} integrates an alternative Huffman-encoding-free lossless pipeline, which is the \LLStage{\bfseries TCMS1-BIT1-RRE1} pipeline with high throughput and decent compression ratio.
In \S\ref{sec:evaluation}, we will see that, {\thiswork} with this lossless pipeline (the throughput-preferred \textbf{{\thiswork}-TP}) also gains very acceptable compression ratios with significantly improved compression throughputs, proving its practical usability.

\subsubsection{The design details of {\thiswork} lossless pipelines}
\begin{figure}
\includegraphics[width=.9\linewidth]{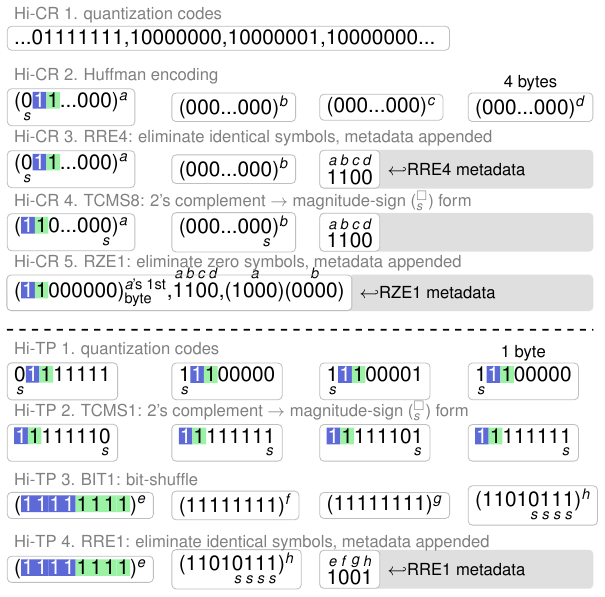}
\caption{Lossless pipelines of {\thiswork}: HF-RRE4-TCMS8-RZE1 (top) and TCMS1-BIT1-RRE1 (bottom). The number in each stage is the width of a single symbol it processes.}
\label{cuszhi::fig::lossless_nohf}
\end{figure}

Figure~\ref{cuszhi::fig::lossless_nohf} illustrates the two selective lossless pipelines employed in {\thiswork} (compressing quantization codes with \LLStage{\bfseries HF-RRE4-TCMS8-RZE1} pipeline or \LLStage{\bfseries TCMS1-BIT1-RRE1} pipeline). The numbers in the module names indicate the per-symbol byte widths. In the \LLStage{\bfseries HF-RRE4-TCMS8} \LLStage{\bfseries -RZE1} pipeline, after the Huffman encoding, \LLStage{RRE4} introduces a bitmap to mark (set to 0) and eliminate symbols identical to their predecessors, subsequently compressing the bitmap recursively. Following this, \LLStage{TCMS8} employs a reversible bitwise operation \begin{center}\verb|(word << 1) ^ (word >> 63)|\end{center} to convert symbols from two's complement to magnitude-sign representation. Finally, \LLStage{RZE1} compresses the symbols in a manner similar to \LLStage{RRE4} but marks symbols equal to zero. The substantial clustering of zeros produced by Huffman encoding, combined with the integration of two reducing stages, maximizes the compression ratio of this pipeline.

The \LLStage{\bfseries TCMS1-BIT1-RRE1} pipeline achieves compression by employing \LLStage{TCMS1} and \LLStage{BIT1} (bit shuffle) for symbol transformation, followed by a single reducing stage, \LLStage{RRE1}. The raw quantization codes remain structured and retain the symmetry of their histogram. Consequently, the symbol widths of the pipeline are uniform, corresponding to the width of a single quantization code. \LLStage{TCMS1} and \LLStage{BIT1} transform the top-1 symbol (128, bit pattern: \texttt{10000000}) and its adjacent symbols (with the most significant bit being \texttt{0} and \texttt{1}, respectively) in such a way that the leading bits of the input stream accumulate a large number of ones. This transformation allows the pipeline to achieve a compression ratio close to that of the previous entropy-plus-spatial encoding pipeline, solely through \LLStage{RRE1}, while significantly enhancing throughput.

\label{sec:lc}

\section{Evaluations}
\label{sec:evaluation}
In this section, we present our evaluations of {\thiswork} on 6 real-world scientific datasets, along with multiple baselines from state-of-the-art GPU-based scientific lossy compressors. From the results, analysis, and discussion, we can clearly find that {\thiswork} achieves optimized compression ratio and quality among all existing works.
\subsection{Experimental setup}
\label{sec:setup}

\subsubsection{Evaluation platforms}

The evaluations are conducted on 2 platforms: one is NERSC-Perlmutter \cite{testbed-nersc}, facilitated with NVIDIA A100 GPUs. The other is a lab-owned workstation with NVIDIA 6000 Ada GPUs. The detailed information on them is in Table \ref{cuszhi::tab::testbed}.

\begin{table}[ht]
  \caption{Testbeds for our experiments.}
  \FloatBodyStyle\footnotesize
  \renewcommand{\arraystretch}{1.2}
  \begin{tabular}{ | >{\bfseries}l|c|c|c|}
    \hline
    {GPU}      & \multicolumn{2}{c|}{A100 (80GB, SXM)}       & RTX 6000 Ada (48 GB)        \\
    \hline
    \hline
    {mem.bw (base-1000)}  & \multicolumn{2}{c|}{2,039 GB/s}    & 960 GB/s  \\
    {mem.bw (base-1024)}  & \multicolumn{2}{c|}{1,900 GiB/s}    & 894 GiB/s \\
    \hline
    \begin{tabular}{@{}c@{}}
      FP32 Performance
    \end{tabular}
    & \multicolumn{2}{c|}{19.5 TFLOPS} & 91.06 TFLOPS \\
    \hline
    CUDA version& \multicolumn{2}{c|}{12.4}        & 12.6        \\
    \hline
    driver version     & \multicolumn{2}{c|}{550.127.08}       & 565.57.01         \\
    \hline
  \end{tabular}%
  \label{cuszhi::tab::testbed}
\end{table}

\subsubsection{Evaluation baselines and configurations}
We include the following GPU-based scientific lossy compressors in the evaluation as baselines: {\cusz}~\cite{cusz,cusz+,cuszi}, cuSZp2~\cite{huang2024cuszp2}, cuZFP~\cite{zfp,cuZFP}, and FZGPU~\cite{FZGPU}. All of them have been widely discussed by the scientific data compression community.
Other existing baselines, such as cuSZx~\cite{szx} and MGARD-GPU~\cite{mgard-latest} are excluded from our evaluation due to significant performance drawbacks shown in past evaluations~\cite{huang2024cuszp2,FZGPU,cuszi}.
To all-roundly present the compression of {\cusz}, its three modes are evaluated, including \textbf{{\cusz}-L} (Lorenzo predictor + Huffman encoder), \textbf{{\cusz}-I} (interpolation predictor + Huffman encoder), and \textbf{{\cusz}-IB} (interpolation predictor + Huffman encoder + NVIDIA Bitcomp). Due to the proprietary and architecture-specific nature of NVIDIA Bitcomp, we evaluate its usage as a separate baseline to compare {\thiswork} with both full-open-source and proprietary solutions. cuSZp2 has 2 modes. To maximize its compression ratio in evaluation, we mostly apply its ``\textbf{outlier} mode'' but will fall back to the more stable ``\textbf{plain} mode'' on use cases when the outlier mode compression exhibits errors.
For our {\thiswork}, as previously introduced in \S\ref{sec:lossless}, we evaluate two modes of it with different lossless pipelines: \textbf{{\thiswork}-CR} and \textbf{{\thiswork}-TP}.

\subsubsection{Evaluation datasets}
{\thiswork} and the baselines are evaluated on 6 real-world scientific datasets (detailed in Table~\ref{tab:datasets}). Those datasets originate from diverse scientific domains and sources~\cite{sdrb,jhtdb,sz3}, and are widely adopted as representative benchmarks for error-bounded scientific lossy compression~\cite {szinterp,SPERR,zfp,HPEZ,cuszi,sdrb}.

\begin{table}[ht]
  \centering
  \caption{Information on the datasets in experiments}
  \FloatBodyStyle\footnotesize
  \renewcommand{\arraystretch}{1.2}
  \begin{tabular}{|rrll|}
    \hline \multicolumn{4}{|l|}{\textbf{CESM-ATM}~\cite{cesm}: Community Earth System Model (\underline{Atm}osphere).}  \\
    & 79 files   & dim: $1800_y\!\times3600_x$          & total: 1.5 GiB \\
    \hline  \multicolumn{4}{|l|}{\textbf{JHTDB}~\cite{jhtdb}: numerical simulation of turbulence.} \\
    & 10 files    & dim: $512_z\times512_y\times512_x$          & total: 5 GiB    \\
    \hline  \multicolumn{4}{|l|}{\textbf{Miranda}~\cite{miranda}: hydrodynamics simulation.} \\
    & 7 files & dim: $256_z\!\times384_y\!\times384_x$          & total: 1 GiB   \\
    \hline
    \multicolumn{4}{|l|}{\textbf{Nyx}~\cite{nyx}: cosmological hydrodynamics simulation.}  \\
    & 6 files & dim: $512_z\!\times512_y\!\times512_x$          & total: 3.1 GiB \\
    \hline
    \multicolumn{4}{|l|}{\textbf{QMCPack}~\cite{qmcpack}: Monte Carlo quantum simulation.}  \\
    & 1 files & dim: $(288\times115)_z\!\times69_y\!\times69_x$ & total: 612 MiB \\
    \hline
    \multicolumn{4}{|l|}{\textbf{RTM}~\cite{geodriveFirstBreak2020}: reverse time migration for seismic imaging.} \\
    & 37 files     & dim: $449_z\!\times449_y\!\times235_x$          & total: 6.5 GiB \\
    \hline

  \end{tabular}
  \label{tab:datasets}
\end{table}

\subsubsection{Evaluation metrics}

Our evaluation metrics are as follows:

\newcommand{\MathCR}{\operatorname{CR}}
\newcommand{\MathSizeof}{\operatorname{sizeof}}

\begin{itemize}

  \item \textbf{Fixed-error-bound compression ratio:}
    We compare the compression ratio (CR) of different compressors on the same data under fixed error bounds. CR is the original input size divided by the compressed size. All error bounds $eb$ in our evaluations are the value-range-based relative error bound, equivalent to a uniform absolute error bound $\epsilon$ (in Eq.~\ref{eq:general}) that equals $eb$ divided by the input data value range.%

  \item \textbf{Rate-distortion:}
    We plot the compression bit rate and the decompression data PSNR for compressors. The bit rate $b$ is the average of bits in the compressed data for each input element (i.e., $32\times$ the reciprocal of CR). PSNR~\cite{z-checker} measures the difference between the input data and decompressed data via value range and mean-square error, and a higher PSNR is better.

  \item \textbf{Throughput:}
    Compression and decompression throughput of all the compressors in GiB/s.

  \item \textbf{Fixed-CR visualization:}
    The visual qualities of reconstructed data from all the compressors at the same CR.

\end{itemize}

\subsection{Evaluation results}

\subsubsection{Compression ratio assessment}
\label{sec:eva-cr}
Table~\ref{cuszhi::tab::CR-trend} presents the compression ratios that {\thiswork} (2 modes, the compression ratio (CR) preferred mode and the throughput (TP) preferred mode) and 5 baselines achieved on all the evaluated datasets with different error bounds (cuZFP is not covered because it doesn't support fixed-error-bound mode).  In almost all test cases, {\cusz}-Hi exhibits the best compression ratio, mainly with the CR-preferred mode. In certain cases under large error bounds, the CR can be extremely high (over 300), and the Huffman tree itself can be a non-negligible overhead in the compressed data, so that the TP-preferred mode without Huffman encoding may have a higher compression ratio than the CR-preferred mode. With synthetic design updates on both the interpolation module and lossless pipeline, {\thiswork} successfully gains the compression ratio improvements over baselines ($>200\%$ for 6/18 cases, $>50\%$ for 11/18 cases), empowering optimized data size reduction over different error-bound constraints. Additionally, it is worth noticing that, as the state-of-the-art compressor in terms of compression ratio, {\cusz}-IB relies on NVIDIA-proprietary components. When compared only with non-proprietary solutions, the open-source {\thiswork} can achieve significantly higher CR improvements, ranging from $113\%$ to $2786\%$ in all test cases.

\begin{table}
  \centering
  \caption{Evaluation of compression ratio (CR). The top 3 CRs per row are shaded in dark blue, light blue, and gray. The last column shows {\thiswork} CR improvement over the baselines.}
  \includegraphics[width=.99\linewidth, trim={0 0 0 0}]{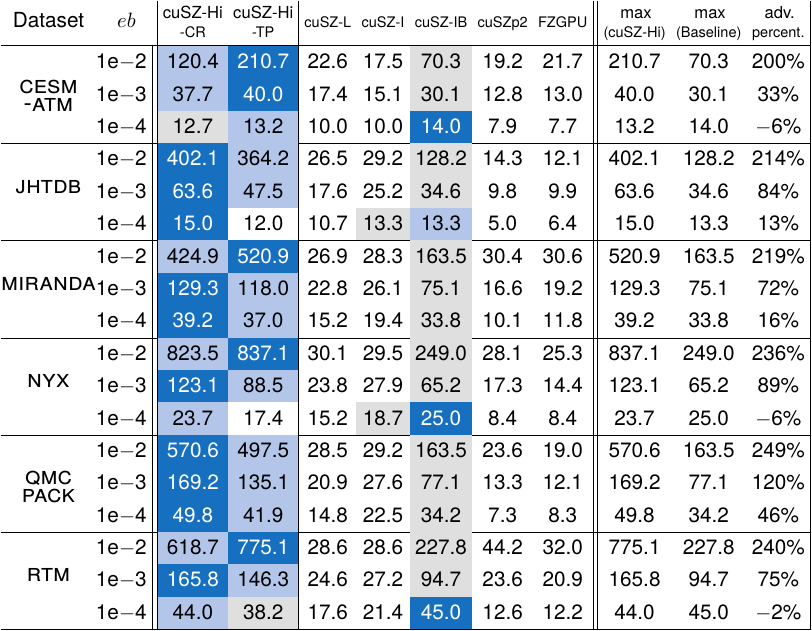}
  \label{cuszhi::tab::CR-trend}
\end{table}

\subsubsection{Rate-distortion assessment}
\label{sec:eva-rd}
\begin{figure}
  \includegraphics[width=\linewidth]{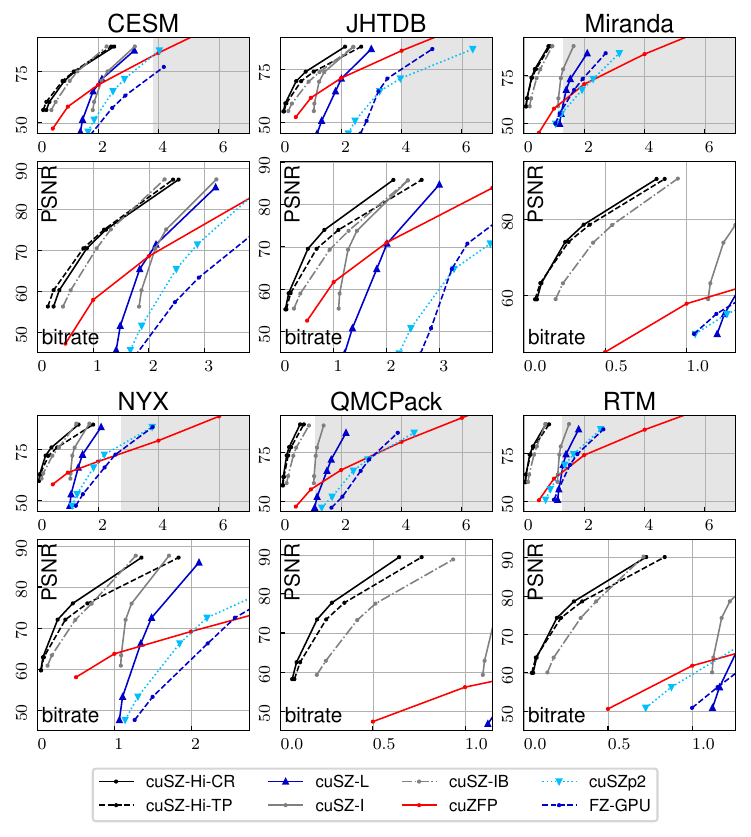}
  \caption{Rate distortion assessment: global (top) and local (bottom) for each dataset. The highly compressible (low-bitrate) regions are indicated by white canvases on top and are zoomed out on the bottom.}
  \label{cuszhi::fig::rate-distortion}
\end{figure}

Besides the compression ratio discussed in \S\ref{sec:eva-cr}, in most scientific lossy compression use cases, the decompression data quality is also critical for justifying the usability of the compression solution. Regarding a joint assessment of compression ratio and data quality, we profile the PSNR metric of decompressed data from different compressors and plot their rate-distortion curves in Figure~\ref{cuszhi::fig::rate-distortion}. For each dataset, the upper rectangle plot displays the full bitrate range for all baselines, and the lower square plot illustrates a narrow bitrate range to visualize better the rate-distortion of the high-ratio compressor {\thiswork} and {\cusz}-I(B). Among all datasets, the CR-preferred mode of {\thiswork} ({\thiswork}-CR) has delivered stable and excellent rate-distortion, achieving the best compression ratio on most data PSNR values. For example, on the Miranda dataset, {\thiswork}-CR achieves $\approx$ $140\%/120\%$ compression ratio improvement over the best baseline {\cusz}-IB when the decompression data PSNR is around 59/63. For compressing the QMCPack dataset, with the decompression data PSNR fixed around 60/76, {\thiswork}-CR achieves $\approx$ $215\%/80\%$ compression ratio improvement over {\cusz}-IB. Moreover, the throughput-preferred mode of {\thiswork} ({\cusz}-Hi-TP) presents excellent rate-distortion as well, exhibiting close decompression data quality to {\thiswork}-CR and outperforming the best baseline ({\cusz}-IB) in many cases. Given the fact that {\cusz}-Hi-TP has a higher throughput than both {\cusz}-IB and {\cusz}-Hi-CR (detailed in \S\ref{sec:eva-tp}), it can become a top compression solution for performance-sensitive use tasks, such as in-time streaming data compression and so on.

\subsubsection{Visualization assessment}
\begin{figure}[ht]
  \footnotesize\fontfamily{qhv}\selectfont
  \begin{tabular}{@{}cc@{}}
    \includegraphics[width=.485\linewidth]{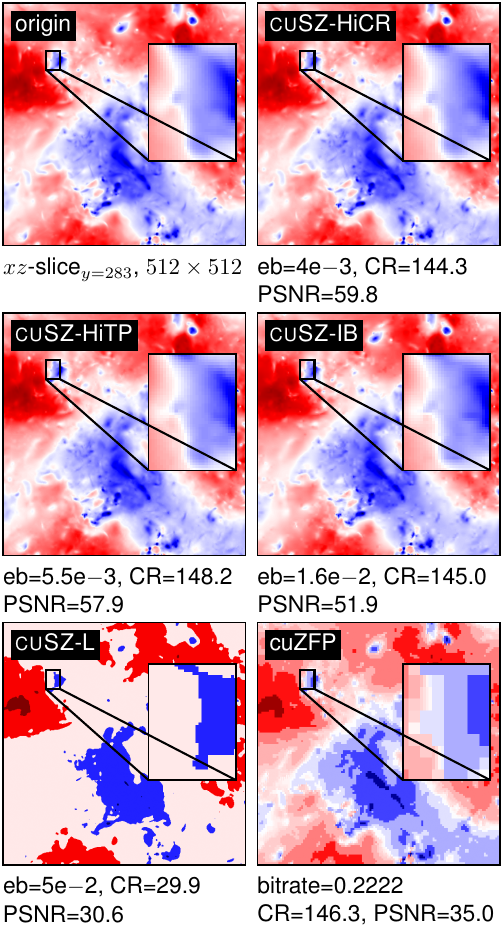} &
    \includegraphics[width=.485\linewidth]{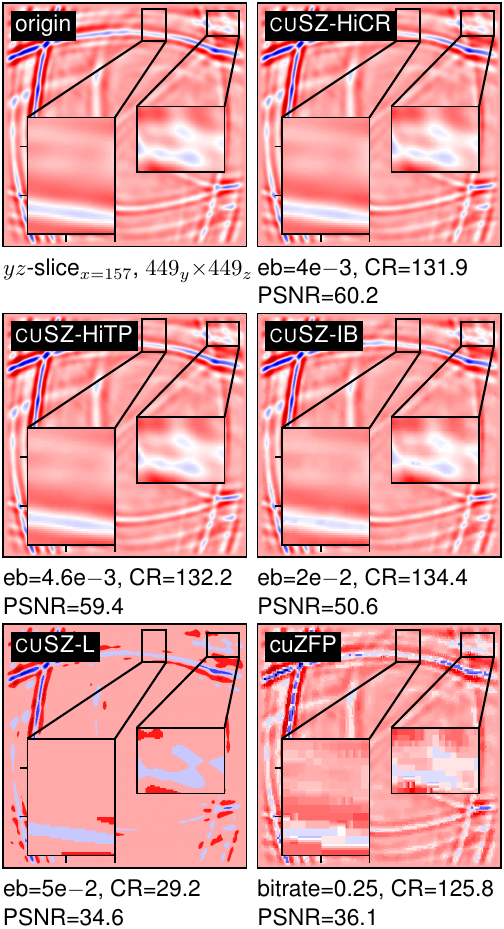} \\
    (a) \bfseries JHTDB \#2500 & (b) \bfseries RTM \#3600
  \end{tabular}
  \vspace{-1\baselineskip}
  \caption{Visualization for quality evaluation.}
  \label{cuszhi::fig::eva-vis}
\end{figure}

Here, we present several detailed showcases of high-quality {\thiswork} compression, visualizing the decompressed data from different compressors at a fixed compression ratio. Figure~\ref{cuszhi::fig::eva-vis} presents 2 series of visualization results on 2 data snapshots (JHTDB-pressure-\#2500 and RTM-\#3600), including the slice visualization of the original data and multiple decompression results from {\thiswork} and baselines (some baselines are omitted from the visualizations as they have very limited compression ratio and/or data quality), with the compression ratios aligned to close values (e.g. $\approx144$ for JHTDB). In each case, in similar compression ratios, when highly noticeable error artifacts appear in the decompression data from baselines ({\cusz}-IB, {\cusz}-L, and cuZFP), the decompression data of {\thiswork} shows the best visualization quality without significant distortions and artifacts from the original data. With the evaluation of both data PSNR and visual quality, we have all-aroundly verified the high compression fidelity of {\thiswork}.

\subsubsection{Speed assessment}
\label{sec:eva-tp}
\begin{figure*}[ht]
  \centering
  \hspace{-5mm}
  \subfigure[\textbf{NVIDIA RTX 6000 Ada}]
  {
    \raisebox{-1cm}{\includegraphics[width=.98\linewidth]{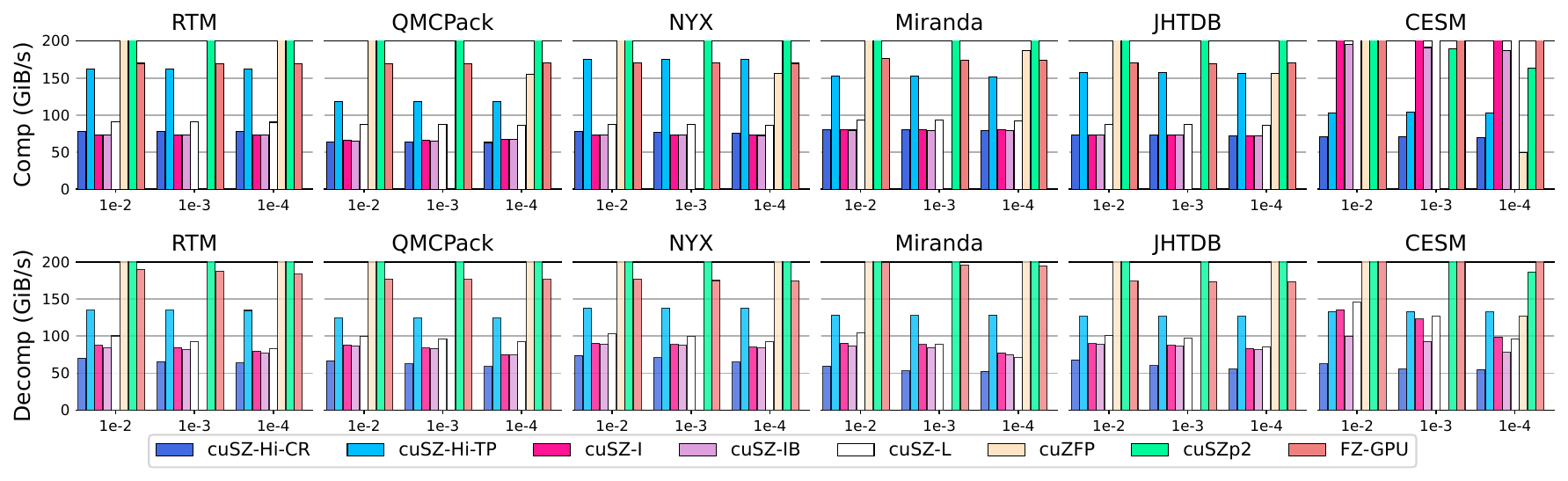}}%
  }
  \hspace{-5mm}

  \vspace{-3mm}
  \hspace{-5mm}
  \subfigure[\textbf{NVIDIA A100}]
  {
    \raisebox{-1cm}{\includegraphics[width=.98\linewidth]{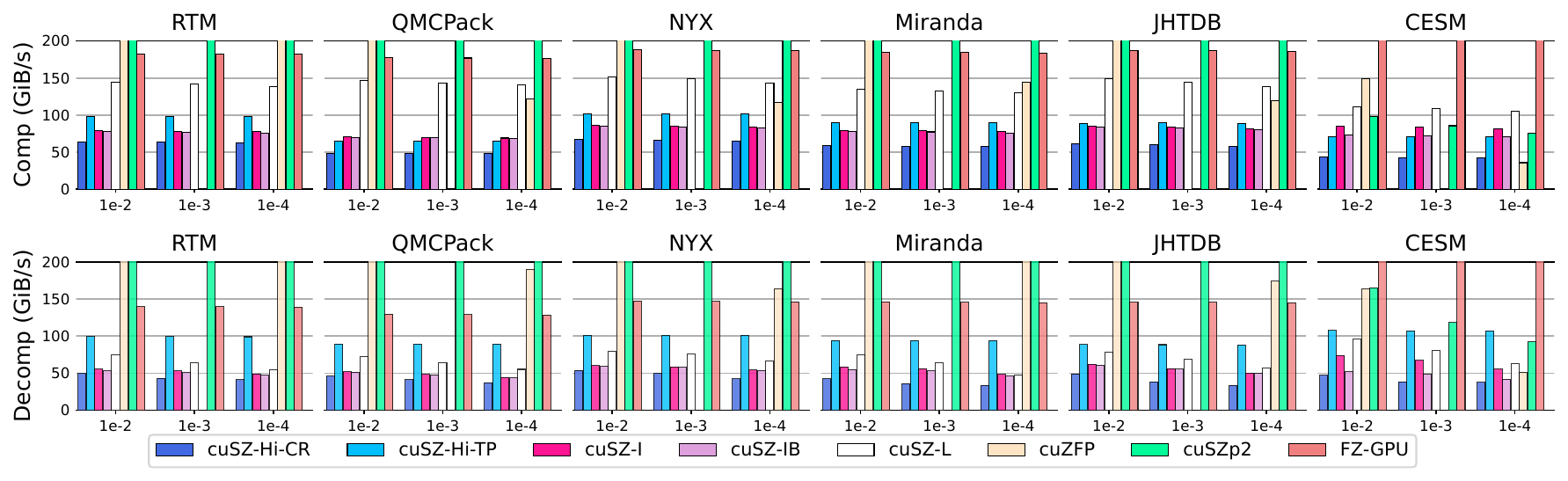}}%
  }
  \hspace{-5mm}
  \vspace{-1mm}
  \caption{Compression and decompression speed in GiB/s on NVIDIA Ada RTX 6000 and A100 with different error bounds. The cuZFP speeds are presented when cuZFP achieves comparable PSNR, or are hidden if no matching PSNR can be found.}
  \label{fig:eva-speed}
\end{figure*}

After verifying the excellent compression ratio and data quality of {\thiswork}, we evaluate its efficiency, proving that {\thiswork} exhibits competitive compression throughput among GPU-based scientific lossy compressors. After profiling the computing throughput (in terms of GPU kernel speed) of several scientific lossy compressors on 2 GPU platforms (NVIDIA A100 and RTX 6000 Ada), we present the results acquired in Figure~\ref{fig:eva-speed}. As cuZFP, cuSZp2, and FZ-GPU are throughput-oriented compressors with limited compression ratios, it is natural that {\thiswork} features relatively lower throughput than them. When we focus the comparison on compressors that can deliver decent compression ratio and quality, we can notice that {\thiswork} has comparable (for CR-preferred mode) or better (for TP-preferred mode) throughput than {\cusz}-L and {\cusz}-I(B). On the NVIDIA RTX 6000 Ada, attributed to its high FP computational power and cache capacity, {\thiswork}-CR shows a similar compression speed and an acceptable $\sim 20\%$ decompression time overhead compared to {\cusz}-I(B) when achieving relatively higher compression ratios. On NVIDIA A100, the compression/decompression overhead of {\thiswork}-CR compared to {\cusz}-I(B) is also mostly limited within $\sim 25\%$. Additionally, while the compression ratio and data quality of {\thiswork}-TP is still better than any baselines, it exhibits a throughput constantly higher than {\cusz}-I(B), even outperforming {\cusz}-L on NVIDIA Ada 6000, showing $\approx 20\% \sim 40\%$ speed improvement over {\cusz}-L (also up to doubled compression speed of {\cusz}-I(B)) on both compression and decompression.
In conclusion, the satisfactory throughput of {\thiswork} successfully guarantees its usability in high-performance scientific computing and data management tasks. Moreover, users can adaptively select either of its two modes, depending on their preference for optimizing compression ratio or computing throughput.

\subsubsection{Ablation study}
At the end of our evaluation, we justify the effectiveness of the newly proposed design components in {\thiswork} with a systematic ablation study. Besides the complete framework of {\thiswork} and the baseline {\cusz}-IB, we also evaluated several incremental frameworks of {\cusz}-IB, stacking separate {\thiswork} features. Those results are reported in Table~\ref{cuszhi::tab::eva-abla}, which includes the compression ratios of the aforementioned compression frameworks on 4 datasets. Specifically, revising the data partition and anchor stride can achieve a compression improvement of $6\% \sim 60\%$ over the original {\cusz}-IB, and further reordering the quantization codes can increase the compression ratio by an additional $6\% \sim 55\%$. Moreover, incorporating auto-tuned selection between multi-dimensional and one-dimensional interpolation into the compression framework increases the compression ratio by $8\% \sim 19\%$. Ultimately, the {\thiswork}-CR with an optimized lossless encoding pipeline achieves a compression ratio that is $16\% \sim 50\%$ higher than before. According to Table~\ref{cuszhi::tab::eva-abla}, each design detail described in \S\ref{sec:details} has contributed well to the improvement of {\thiswork} compression ratio.

\begin{table}[ht]
  \centering
  \caption{Ablation study: compression ratios of {\cusz}-IB
  , {\thiswork}, and different design increments between them.}
  \includegraphics[width=0.99\linewidth]{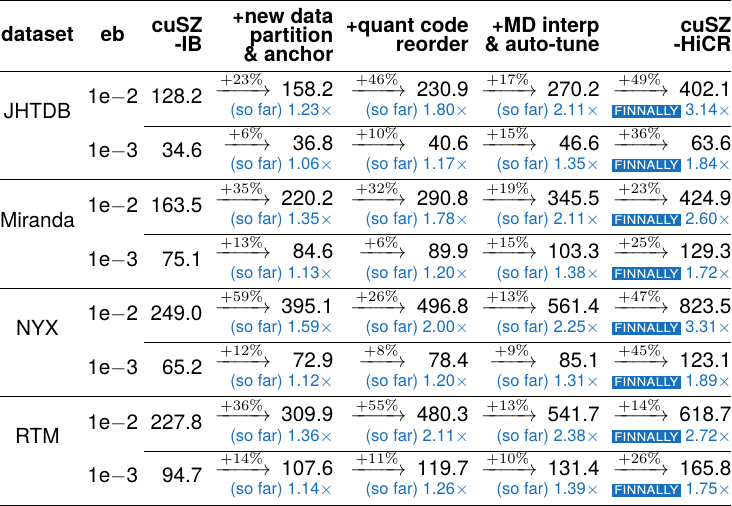}
  \label{cuszhi::tab::eva-abla}
\end{table}

\section{Conclusion}\label{sec:conclusion}

Scientific error-bounded lossy compression is an essential technique for exascale scientific data management. Unfortunately, as fast-evolving GPUs play an increasingly important role in scientific supercomputing, existing scientific error-bounded lossy compression toolkits on GPU platforms remain immature and suboptimal. This paper proposes {\thiswork}, the first GPU-based high-ratio, high-quality, and open-source scientific error-bounded lossy compressor. Integrating highly advanced and GPU-customized data interpolation schemes and also fine-tuned lossless encoding pipelines, {\thiswork} significantly outperforms state-of-the-art GPU-based scientific lossy compressors. When exhibiting close to or better throughput than the state-of-the-art high-ratio GPU-based compressor {\cusz}-I, which utilizes an NVIDIA-proprietary encoding module, {\thiswork} can significantly improve its compression ratio by up to $ 200\%$ in the same compression task.
In the future, we will improve {\thiswork} in the following aspects: 1) Reduce the interpolation computing overhead on GPU platforms; 2) Explore and design better lossless encoders for quantization codes; 3) Enhance the flexibility of compression, such as develop an auto-selection mechanism for different data compressor archetypes and/or lossless pipelines to fit different data characteristics and compression requirements dynamically.

\begin{acks}
This research was supported by the U.S. Department of Energy, Office of Science, Advanced Scientific Computing Research (ASCR), under contracts \texttt{DE-AC02-06CH11357}. This work was also supported by the National Science Foundation (Grant Nos. \texttt{2104023}, \texttt{2311875}, \texttt{2344717}, \texttt{2514034}, \texttt{2514035}, and \texttt{2514036}).
\end{acks}

\newpage

\bibliographystyle{ACM-Reference-Format}
\bibliography{references}
\end{document}